# Superconductivity at 215 K in lanthanum hydride at high pressures

A. P. Drozdov, V. S. Minkov, S. P. Besedin, P. P. Kong, M. A. Kuzovnikov, D. A. Knyazev, M. I. Eremets

*Max-Planck-Institut fur Chemie, Hahn-Meitner Weg 1, 55128 Mainz, Germany*

*We synthesized lanthanum hydride (LaHx) by laser heating of lanthanum in hydrogen atmosphere at pressure P = 170 GPa. The sample shows a superconducting step at ~209 K and 170 GPa and temperature dependence of resistance. By releasing the pressure to ~150 GPa, the superconducting transition temperature $T_c$ increases to ~215 K – the record $T_c$. This finding supports a way of achieving $T_c$ higher than the one in $H_3S$ (203 K)[1] – in hydrides with sodalite-like structures, first proposed for $CaH_6$[2] ($T_c$=245 K) and later for yttrium and lanthanum hydrides where higher, room temperature superconductivity is expected[3,4].*

The current record of $T_c$=203 K has been established in $H_3S$[1]. That was the first confirmation of the prediction of Migdal-Eliashberg theory of possibility of unlimitedly high temperature superconductivity, with phonon-mediated pairing, and opening a way for the room temperature superconductivity. $T_c$ can be high at significantly high characteristic phonon energies and electron-phonon interaction. Hydrogen provides the highest phonon frequencies and therefore it is naturally to expect high $T_c$ in hydrogen materials. An ultimate case is metallic hydrogen where very high critical temperature for superconductivity $T_c$ =200-400 K was predicted at P>400-500 GPa[5,6]. Superconducting hydrogen has a great fundamental interest but achieving these high pressures is a long-standing challenge. Therefore, the next prediction of Ashcroft[2] brought up greater interest: he proposed to search for superconductivity in hydrogen-dominated materials ($CH_4$, $SiH_4$, $GeH_4$ etc.). These materials transform to metals at much lower, accessible pressures and could be high temperature superconductors (HTSC) for reasons similar to pure metallic hydrogen: high values of characteristic phonon frequencies and strong electron-phonon coupling. Ashcroft's paper[7] contains a qualitative consideration and is too general to guide experiments. It turned out that some of the proposed hydrides dissociate or remains insulators; for instance, methane at pressures about 520 GPa[8]. At the same time, methods of crystal structure prediction have been developed[9-12]. Since then, many electron and phonon spectra has been calculated for different hydrides[13-15]. High superconducting transition temperatures, estimated from the McMillan formula of the BCS theory or more precisely from solution of the Eliasberg equations, are persistently predicted ($T_c$>100 K).

Recently, almost all binary hydrides have been studied[13-15], and the study of the ternary compounds are on the way. $T_c$ values greater than 200 K, and even room temperature superconductivity is predicted in the family of hydrides: $T_c$~320 K at 250 GPa[3] (or 303 K at 400 GPa according to Ref.[4]) for $YH_{10}$; and $T_c$~280 K at ~200 GPa for $LaH_{10}$ [3,4]. All these compounds have clathrate structure: hydrogen atoms are connected to each other with weak covalent bonds and with ionic bonding to the host atom in the center of the sodalite structure. This structure is different from the *Im-3m* structure of $H_3S$, where strong covalent bonds attach sulphur atoms to hydrogen atoms.

The sodalite structure for calcium hydride with $T_c$~235 K was predicted some time ago[19], and in our group we immediately started to work on this hydride and have been continuing the study, but it turned out to be a difficult problem, and we succeeded in finding superconductivity in $H_3S$ quicker[1]. Experimental verification of the predicted superconductivity in $CaH_6$ and other hydrides with the clathrate structure is quite difficult, because the sample should be synthesized at very high pressures ~200 GPa, by heating a metal in hydrogen atmosphere at temperature T~1000 K. Characterization of the samples is also challenging, especially electrical measurements which are indispensable for evidencing of superconductivity. Recently, significant progress was achieved by R. Hemley group[20], where lanthanum hydride was synthesized at P>160 GPa and T~1000 K. The X-ray data indicate that the stoichiometry should be LaH$_{10\pm x}$ (x is between +2 and -1.) – close to LaH$_{10}$. To our knowledge no results on superconductivity has published so far. Recently, at a conference[21], the authors demonstrated a broad step at the temperature dependence of resistance, which can be a sign of a superconducting transition at ~260 K (resistance, however, did not drop to zero upon cooling the sample).

Following our long-term efforts to find superconductivity in Ca, Y and La hydrides we finally succeeded in finding superconductivity in lanthanum hydride. We loaded a sample of La (Alfa Aesar 99.9%) in a hydrogen atmosphere in a gas loader at pressure at P~0.1 GPa. DAC with anvils with 60 µm culet were used. The gasket which was made of T301 steel is separated from the electrical leads (Ta covered with Au, sputtered on the anvil) by an insulating layer of made of CaF$_2$ mixed with epoxy. After clamping the sample in hydrogen atmosphere at pressure ~2 GPa, the DAC was extracted and the pressure was further increased at room temperature to P~170 GPa. Figure 1 shows photographs of the sample inside DAC at P~171 GPa. Photos are taken in combined transmission-reflection illumination. Raman spectra of the lanthanum hydride at different pressures are shown in Fig. 2. Only the hydrogen vibron was observed in the Raman spectra. At P>146 GPa new low frequency peaks appeared, the most prominent one at ~800 cm$^{-1}$. The Raman spectra evidence that metallic La transformed to a non-metallic hydride LaH3[1]. Significant decrease of intensity of hydrogen vibron (measured over the sample at the same position) also indicates reaction of La with hydrogen.

At 170 GPa, we heated the sample with a YAG laser at temperatures below ~1000 K. As a result, the volume of the sample was significantly increased while hydrogen deceased. Hydrogen still remained around the sample. Figure 3 shows the temperature dependence R(T) of the resistance of the heated sample. A pronounced step indicating a superconducting transition can clearly been seen in Fig 3. The heated sample was in a disordered state, as can be judged from the specific shape of the superconducting step: there is a pronounced peak at the onset and a step on the transition to superconducting state – features which are well known for disordered superconductors[22]. These features (the peak and the steps at the transition to the superconductivity) disappeared after further annealing with the aid of the next laser heating, which could be explained by an improvement in the crystallinity. T$_c$ only slightly increased with the annealing.

By decreasing the pressure to 150 GPa, T$_c$ increased to the record value of ~215 K. Further decreasing of the pressure to 138 GPa, decreases the T$_c$ substantially. At the next cooling at this pressure the superconducting state step disappears (Fig. 3). Interestingly, by increasing the pressure to P~160 GPa again, superconductivity re-appears (Fig. 3). However, the shape of the SC step (steps at the slope and the peak at the R(T) curve) indicate that the sample is disordered after being back to superconducting state[22]. The disappearance of superconductivity by decreasing the pressures below P~140-150 GPa and its recovering with the pressure increase indicate on a structural transition. Likely this is a structural phase transformation because the superconductivity does not recover reversibly at the back transformation. This phase transformation likely is the same as observed in Ref. [20] at 150 GPa by X-ray diffraction studies.

This work present the first observation of a superconductivity with a record T$_c$ in a new class of hydrides which are predicted to have a clathrate structure[2]. Experiments for further proofs such as dependence the superconducting transition, magnetic field, Meissner effect, isotope effect, and X-ray diffraction are in progress.

Acknowledgements. Support provided by the European Research Council under Advanced Grant 267777 is acknowledged. The authors appreciate Sh. Mozaffari for helpful comments. ME is thankful the Max Planck community for the invaluable support, and U. Pöschl for the constant encouragement.

(a)    170 GPa before laser heating

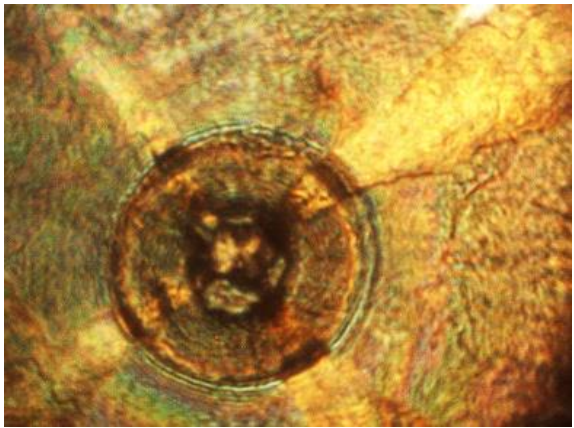 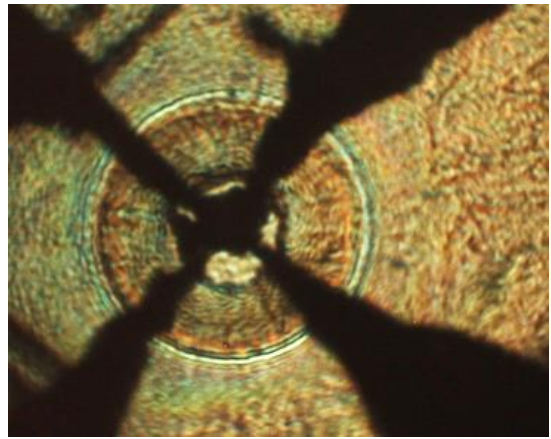

(b)    170 GPa after laser heating

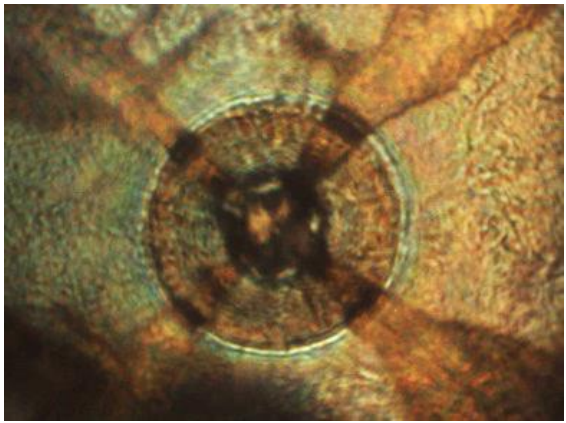 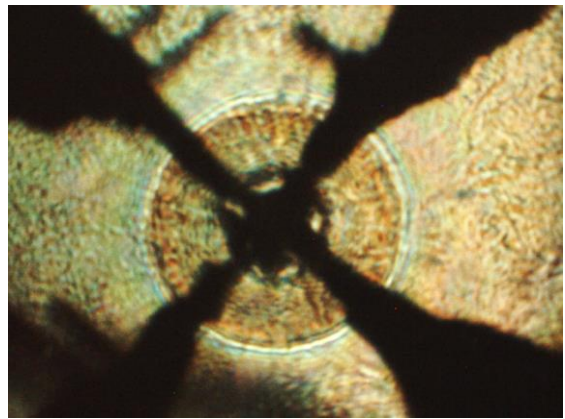

Fig. 1. View of the sample inside DAC at 171 GPa with the attached four electrodes.
(a) Left – photo is taken in a combined transmission- reflection illumination, and right –in transmission illumination.
(b) The laser heated the sample at the same pressure – it is significantly increased in volume and filled nearly the whole sample space cavity. The sample is still surrounded by hydrogen.

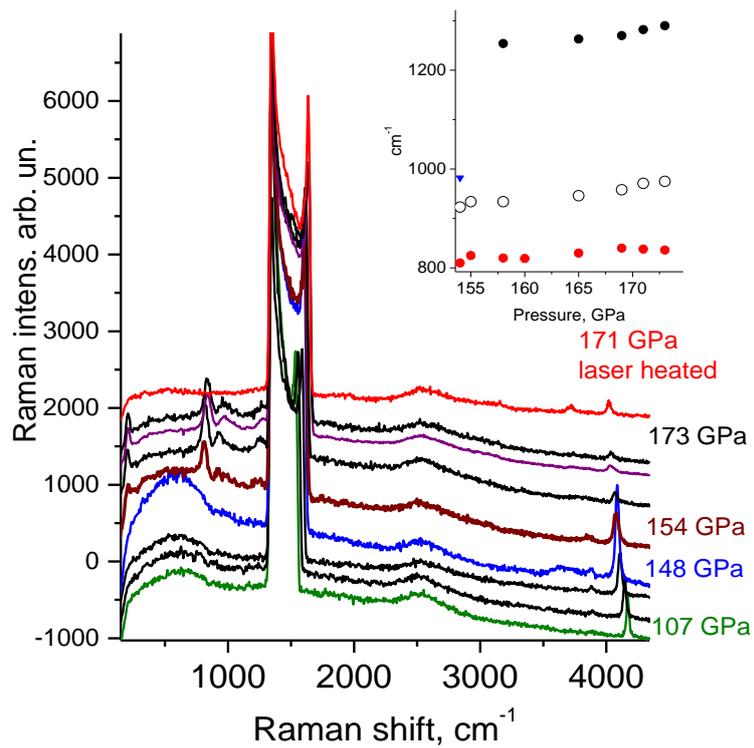

Fig. 2. Raman spectra of the lanthanum hydride at different pressures, taken from the point indicated by an arrow in Fig. 1a. The appearance of a new peaks at ~800 cm$^{-1}$ indicate creation of an insulating phase of LaH$_x$ at pressures above ~150 GPa [23]. Inset shows the pressure dependence of the Raman peaks. These peaks disappear after laser heating at 171 GPa, indicating a transformation to a metal.

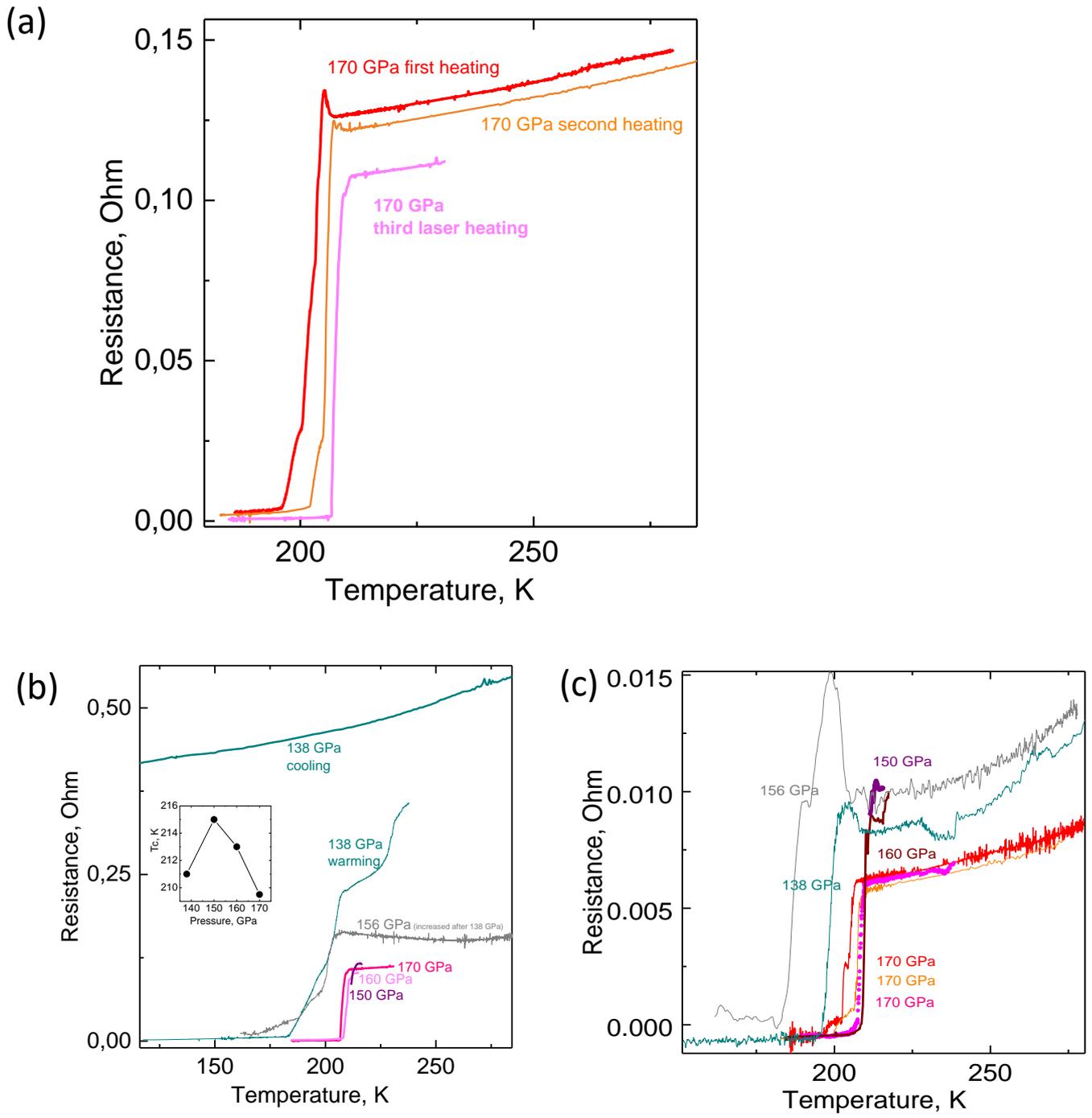

Fig. 3. Electrical measurements of the lanthanum hydride sample heated at 170 GPa in hydrogen atmosphere with the aid of a laser. The presented measurements were done during slow (~1 K/min) warming.
(a) Data from one channel of four-probe van der Pauw measurements. A superconductive step appeared after the first laser heating (red curve). The peak at the onset of the superconducting transition, the broadening of the transition, and the oscillations at the slope of superconducting transition indicate a disordered state for the sample as it is known from literature [22]. The sample is subsequently annealed at the same pressure with successive heating as follows from the change of the shape of the superconducting.

(b) By decreasing the pressure, $T_c$ increases, but drops at pressures below 150 GPa. Inset shows the pressure dependence of $T_c$. The measurements at 138 GPa show a broad superconducting transition. Over time, superconductivity at this pressure (138 GPa) disappeared (see 138 GPa cooling). Superconductivity reappeared when pressure was increased to 156 GPa.

(c) Data from another channel of the van der Pauw measurements. For disordered state of the sample (measurements at 138 GPa and 156 GPa) the shape of the superconducting transition is different from the data from another channel (b, c).

**References**

1   Drozdov, A. P., Eremets, M. I., Troyan, I. A., Ksenofontov, V. & Shylin, S. I. Conventional superconductivity at 203 K at high pressures. *Nature* **525**, 73-77, doi:doi:10.1038/nature14964 (2015).
2   Wang, H., Tse, J. S., Tanaka, K., Iitaka, T. & Ma, Y. Superconductive sodalite-like clathrate calcium hydride at high pressures. *PNAS* **109**, 6463–6466 (2012).
3   Liu, H., Naumov, I. I., Hoffmann, R., Ashcroft, N. W. & Hemley, R. J. Potential high-Tc superconducting lanthanum and yttrium hydrides at high pressure. *PNAS*, doi/10.1073/pnas.1704505114 (2017).
4   Peng, F. *et al.* Hydrogen Clathrate Structures in Rare Earth Hydrides at High Pressures: Possible Route to Room-Temperature Superconductivity. *PRL* **119** 107001 (2017).
5   Ashcroft, N. W. Metallic hydrogen: A high-temperature superconductor? *Phys. Rev. Lett.* **21**, 1748-1750 (1968).
6   Borinaga, M., Errea, I., Calandra, M., Mauri, F. & Bergara, A. Anharmonic effects in atomic hydrogen: Superconductivity and lattice dynamical stability. *Phys. Rev. B* **93**, 174308 (2016).
7   Ashcroft, N. W. Hydrogen Dominant Metallic Alloys: High Temperature Superconductors? *Phys. Rev. Lett.* **92**, 187002 (2004).
8   Gao, G. *et al.* Dissociation of methane under high pressure. *J. Chem. Phys.* **133** 144508 (2010).
9   Pickard, C. J. & Needs, R. J. Ab initio random structure searching. *J. Phys.: Condens. Matter* **23**, 053201 (2011).
10  Y.Wang, Lv, J., Zhu, L. & Ma, Y. Crystal structure prediction via particle-swarm optimization. *Phys. Rev. B* **82**, 094116 (2010).
11  Oganov, A. R. & Glass, C. W. Crystal structure prediction using evolutionary algorithms: Principles and applications. *J. Chem. Phys.* **124**, 244704 (2006).
12  Zurek, E. in *Handbook of Solid State Chemistry, 1* (ed Shinichi Kikkawa Richard Dronskowski, and Andreas Stein.) Ch. 15, 571-605 (Wiley-VCH Verlag GmbH & Co. KGaA, 2017).
13  Zhang, L., Wang, Y., Lv, J. & Ma, Y. Materials discovery at high pressures. *Nature Rev. Mater.* **2**, 17005 (2017).
14  Wang, H., Li, X., Gao, G., Li, Y. & Ma, Y. Hydrogen-rich superconductors at high pressures. *WIREs Comput Mol Sci* **e1330**, doi:doi: 10.1002/wcms.1330 (2017).
15  Bi, T., Zarifi, N., Terpstra, T. & Zurek, E. The Search for Superconductivity in High Pressure Hydrides. *arXiv preprint arXiv:1806.00163* (2018).

16  Feng, J. e. a. Structures and potential superconductivity in SiH4 at high pressure: en route to ''metallic hydrogen *Phys. Rev. Lett.* **96** 017006 (2006).
17  Eremets, M. I., Trojan, I. A., Medvedev, S. A., Tse, J. S. & Yao, Y. Superconductivity in Hydrogen Dominant Materials: Silane. *Science* **319**, 1506-1509 (2008).
18  Goncharenko, I. *et al.* Pressure-Induced Hydrogen-Dominant Metallic State in Aluminum Hydride. *Phys. Rev. Lett.* **100**, 045504 (2008).
19  Wang, H., Tse, J. S., Tanaka, K., Iitaka, T. & Ma, Y. Superconductive sodalite-like clathrate calcium hydride at high pressures. *PNAS* **109**, 6463–6466 (2011).
20  Geballe, Z. M. *et al.* Synthesis and Stability of Lanthanum Superhydrides. *Angew. Chem. Int. Ed.* **57**, 688 –692 (2018).
21  Symposium organized by Areces Foundation "Superconductivity and Pressure, a Fruitful Relationship: Towards Room Temperature superconductivity" *Madrid 20-23rd of May 2018* (2018).
22.  Zhang, G. *et al.* Bosonic Anomalies in Boron-Doped Polycrystalline Diamond. *Phys. Rev. Appl.* **6**, 064011 (20161 Meng, H., Kuzovnikov, M. A. & Tkacz, M. Phase stability of some rare earth trihydrides under high pressure. *Int. J. Hydrogen Energy* **42**, 29344-29349 (2017).
23.  Meng, H., Kuzovnikov, M. A. & Tkacz, M. Phase stability of some rare earth trihydrides under high pressure. *Int. J. Hydrogen Energy* **42**, 29344-29349 (2017).